\documentclass[lettersize,journal]{IEEEtran}
\usepackage{amsmath,amsfonts}
\usepackage{algorithmic}
\usepackage{algorithm}
\usepackage{array}
\usepackage[caption=false,font=normalsize,labelfont=sf,textfont=sf]{subfig}
\usepackage{textcomp}
\usepackage{stfloats}
\usepackage{url}
\usepackage{verbatim}
\usepackage{graphicx}
\usepackage{cite}
\usepackage{siunitx}
\usepackage{amsmath,amssymb,amsfonts}
\usepackage{algorithmic}
\usepackage{wrapfig}
\usepackage{booktabs}
\usepackage{threeparttable}

\begin{document}

\title{Coarse-to-Fine Non-Rigid Registration for Side-Scan Sonar Mosaicking}

\author{Can Lei,
        Nuno Gracias,
        Rafael Garcia,
        Hayat Rajani,~\IEEEmembership{Member,~IEEE,}
        and Huigang Wang,~\IEEEmembership{Member,~IEEE}
\thanks{Can Lei and Huigang Wang are with the School of Marine Science and Technology, Northwestern Polytechnical University, Xi’an 710072, China. Huigang Wang is also with the Research \& Development Institute of Northwestern Polytechnical University in Shenzhen, Shenzhen 518057, China.}

\thanks{Hayat Rajani, Nuno Gracias and Rafael Garcia are with the Computer Vision and Robotics Research Institute (ViCOROB) of the University of Girona, Spain. This work was conducted while Can Lei was on a research stay at ViCOROB, Spain}

\thanks{This work was partly supported by the Spanish government through projects ASSiST (PID2023-149413OB-I00) and IURBI (CNS2023-144688). This work was also supported by the National Natural Science Foundation of China (62171368) and Science, Technology and Innovation of Shenzhen Municipality (KJZD20230923115505011, JCYJ20241202124931042).}
\thanks{Corresponding author: Huigang Wang (e-mail: wanghg74@nwpu.edu.cn).}}



\maketitle

\begin{abstract}
Side-scan sonar mosaicking plays a crucial role in large-scale seabed mapping but is challenged by complex non-linear, spatially varying distortions due to diverse sonar acquisition conditions. Existing rigid or affine registration methods fail to model such complex deformations, whereas traditional non-rigid techniques tend to overfit and lack robustness in sparse-texture sonar data. To address these challenges, we propose a coarse-to-fine hierarchical non-rigid registration framework tailored for large-scale side-scan sonar images. Our method begins with a global Thin Plate Spline initialization from sparse correspondences, followed by superpixel-guided segmentation that partitions the image into structurally consistent patches preserving terrain integrity. Each patch is then refined by a pretrained SynthMorph network in an unsupervised manner, enabling dense and flexible alignment without task-specific training. Finally, a fusion strategy integrates both global and local deformations into a smooth, unified deformation field. Extensive quantitative and visual evaluations  demonstrate that our approach significantly outperforms state-of-the-art rigid, classical non-rigid, and learning-based methods in accuracy, structural consistency, and deformation smoothness on the challenging sonar dataset.

\end{abstract}


\section{Introduction}
\IEEEPARstart{S}{ide}-scan sonar (SSS) is widely used for large-scale seabed mapping \cite{A1} due to its wide swath and high-resolution imaging. To generate coherent sonar mosaics, multiple SSS images acquired from different viewpoints or times need to be accurately aligned. Image registration thus plays a vital role in this process, enabling downstream tasks such as target detection \cite{A2} and seafloor classification \cite{A3}. While rigid or affine models, typically based on keypoint matching and RANSAC, are commonly used \cite{A4,A5}, they struggle with non-linear, spatially varying distortions caused by platform motion and seafloor geometry. These challenges highlight the need for non-rigid registration methods tailored to the unique characteristics of sonar imagery.

Non-rigid registration of SSS imagery presents unique challenges rooted in the physics of sonar imaging. The same seafloor structures can appear markedly different across overlapping strips due to changes in viewing geometry, including variations in platform trajectory, heading, and incidence angle, resulting in complex, spatially non-uniform distortions \cite{A6}. In addition, sonar images are degraded by low texture diversity, noise, and uneven intensity, which interfere with feature extraction and matching \cite{A7}. These factors complicate the establishment of stable correspondences, particularly in regions lacking distinctive structures or corrupted by noise. Therefore, effective registration must not only align large-scale structural layouts, but also remain robust to local ambiguities and structural inconsistencies necessitating tailored strategies that integrate both global guidance and locally adaptive alignment mechanisms.

To address the above challenges, we introduce a coarse-to-fine non-rigid registration framework tailored for large-scale SSS mosaicking. Unlike prior approaches that rely solely on traditional models or task-specific supervision, our method incorporates structural guidance, noise robustness, and hierarchical alignment strategies. The key contributions of this work are as follows:

\begin{itemize}
    \item We formulate a hierarchical registration paradigm that integrates global structure-aware initialization with locally adaptive refinement, enabling accurate alignment under complex and spatially variant distortions.

    \item We introduce a superpixel-assisted patch extraction scheme that segments the SSS image into rectangular patches guided by irregular superpixel regions, preserving local structural integrity and providing context-rich inputs for effective local registration.

    \item We apply a pretrained SynthMorph-based unsupervised registration module to each local patch pair, enabling dense and flexible alignment without any task-specific training.
    
\end{itemize}

\section{Related Work}

Based on the transformation model complexity, existing registration approaches for SSS image are typically classified into four categories: rigid \cite{A8}, affine \cite{A9}, projective \cite{A10}, and non-rigid registration \cite{A11}.

Rigid registration assumes that the geometric relationship between two SSS images can be modeled by a combination of rotation and translation \cite{A12}. Such methods offer high computational efficiency and are relatively straightforward to implement \cite{A13,A14}. However, they are inherently sensitive to variations in viewpoint, scale, and orientation. Consequently, rigid registration is generally applicable only when platform motion is stable and seafloor structures remain unchanged. For instance, Li et al. \cite{A15} employed region-based feature extraction for sonar image matching, followed by rigid transformation to estimate the rotation matrix and translation vector for image alignment. To enhance robustness, Vandrish et al. \cite{A16} utilized the Scale-Invariant Feature Transform (SIFT) to extract and match keypoints, and applied Random Sample Consensus (RANSAC) for outlier rejection and rigid transformation estimation. Despite their simplicity, rigid methods fail to accommodate scale discrepancies and local deformations frequently observed in sonar imagery due to platform instability and seafloor undulations.

To address more complex geometric variations, affine registration introduces additional degrees of freedom, such as scaling and shearing, thereby enabling more flexible alignment \cite{A17,A18,A19}. For instance, Zhang et al. \cite{A20} extracted and matched Curvelet-based features between adjacent sonar swaths and applied affine transformation for alignment. Tao et al. \cite{A21} incorporated navigational priors to guide the extraction of affine-invariant SURF features, followed by affine estimation via RANSAC, leading to improved alignment accuracy. Although affine models offer increased tolerance to linear distortions compared to rigid models, they still operate under a linear transformation assumption and often fail under conditions involving perspective effects or pronounced local deformation.

Projective registration, typically formulated via homography estimation \cite{A22,A23}, further generalizes the transformation model to account for perspective distortion \cite{A24}, making it more applicable in SSS scenarios with large viewpoint shifts or complex seafloor geometry. Wang et al. \cite{A25} used AKAZE features and homography estimation via RANSAC to register SSS images, while Tang et al. \cite{A26} proposed a SURF-based mosaicking framework involving robust matching and global homography alignment. These approaches improve robustness against viewpoint changes and enhance visual continuity. However, they assume global planarity, limiting their effectiveness in modeling spatially localized nonlinear distortions.

To better accommodate complex and spatially non-uniform deformations, non-rigid registration techniques have been extensively explored \cite{A27}. These include parametric models such as Thin Plate Splines (TPS) \cite{A28} and B-splines \cite{A29}, as well as non-parametric dense deformation methods like optical flow \cite{A30} and the Demons algorithm \cite{A31}. For example, Shang et al. \cite{A32}integrated Coordinates and trackline constraints with SURF-based matching to identify overlapping regions, followed by TPS to capture spatially varying distortions. Da et al. \cite{A33} introduced a Demons-based registration framework for forward-looking sonar sequences, where structural features derived from gradient orientations and magnitudes guided the deformation field estimation. While traditional non-rigid methods provide smooth deformation models, they are often prone to overfitting, lack robustness in texture-sparse sonar imagery, and typically fail to incorporate structural constraints critical for reliable registration.

In recent years, deep learning has emerged as a powerful alternative for estimating dense deformation fields, especially in the context of medical or multimodal image registration \cite{A34,A35}. For instance, Fu et al. \cite{A36} proposed a LoFTR-based deep matching framework to register multibeam and side-scan sonar images, utilizing attention mechanisms to overcome cross-modal inconsistencies and geometric distortion. Despite their robustness, deep learning-based registration methods require large-scale annotated datasets for supervised training. However, such datasets are unavailable in sonar applications, thereby limiting the practicality of these approaches in underwater scenarios.

In summary, traditional non-rigid methods provide smooth deformation priors but exhibit limited generalization and structural adaptability. Deep learning-based approaches demonstrate strong potential but are constrained by data availability and training requirements. Furthermore, most existing frameworks rely on global deformation models and thus lack the capacity to adapt to local geometric variations across diverse seafloor environments. These limitations motivate the development of hybrid non-rigid registration strategies that can jointly leverage global priors and local adaptability, particularly for the low-texture, locally distorted conditions commonly encountered in SSS mosaicking.

\section{Method}

\subsection{Approach Overview}

\begin{figure*}[htbp]
	\centering
	\includegraphics[width=1\linewidth]{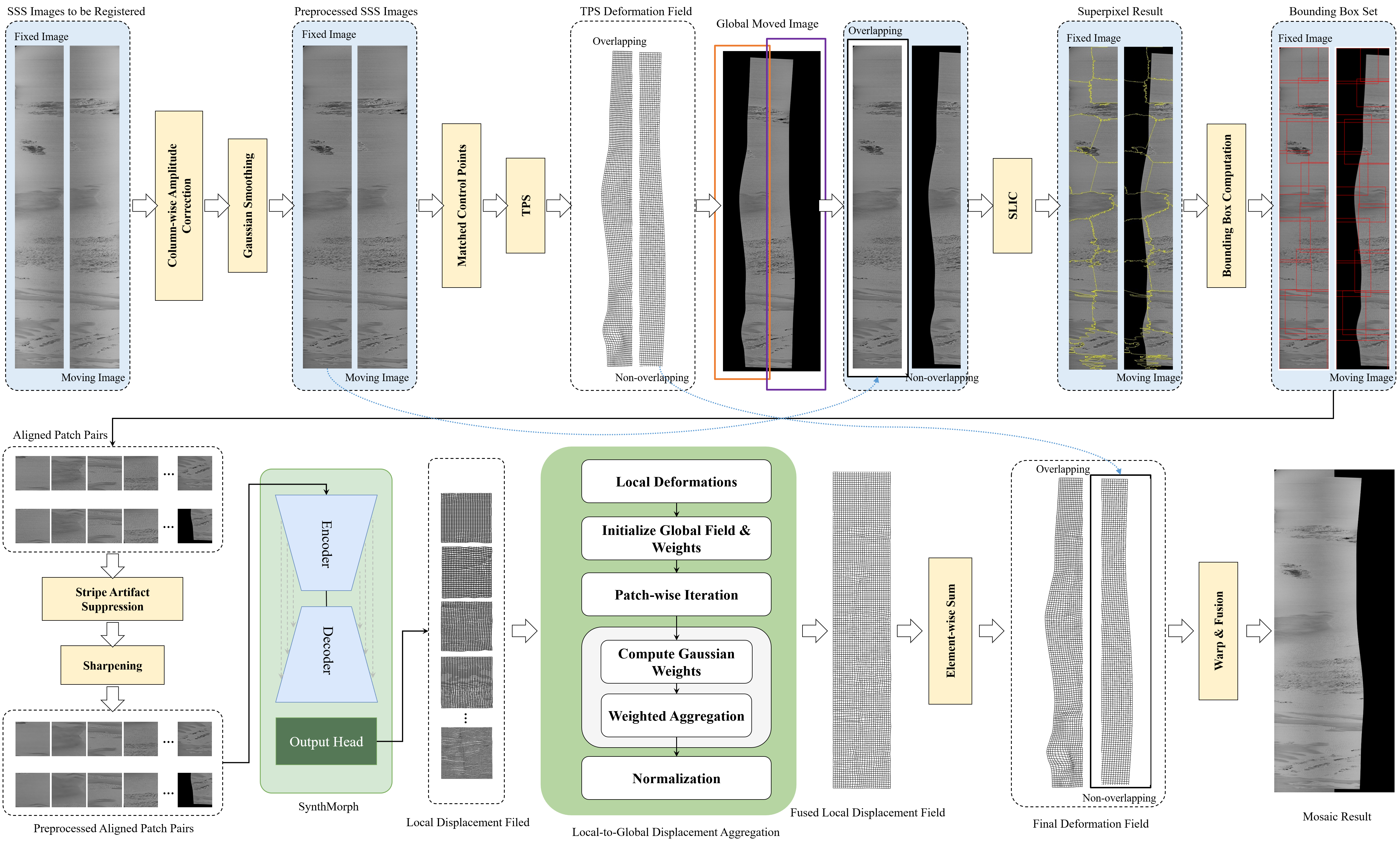}
	\caption{Coarse-to-fine hierarchical registration framework for side-scan sonar images. Global TPS warping compensates large-scale distortions, followed by superpixel-based local partitioning. Each local patch undergoes artifact suppression preprocessing and unsupervised deep registration to estimate dense deformation fields. Local deformations are fused into a global transformation, combining TPS-based global modeling and deep learning-based local refinement for accurate non-rigid alignment.}
	\label{fig00}
\end{figure*}	

To tackle complex spatial distortions and non-rigid deformations in side-scan sonar images, we propose a coarse-to-fine hierarchical registration framework. As illustrated in Fig. \ref{fig00}, the method first applies a global geometric correction based on TPS warping to compensate for large-scale structural displacements, providing a reliable initialization for further refinement. Subsequently, the image is partitioned into structurally consistent local regions using superpixel segmentation, with each region encapsulated by its minimum bounding rectangle to standardize patch shapes. Within each local patch, artifact suppression preprocessing is performed before applying a pretrained, unsupervised deep registration network to estimate dense deformation fields, enabling accurate local alignment. Finally, all local deformation fields are aggregated into a unified global transformation. This modular design effectively combines classical geometric modeling and modern learning-based techniques, enhancing registration accuracy and robustness without relying on task-specific training.

\subsection{Initial Global Non-Rigid Alignment}

In SSS images, range-dependent acoustic attenuation causes intensity inhomogeneity across the swath. To mitigate this effect and achieve an initial non-rigid alignment (Fig.~\ref{fig01}), we first perform swath-wise radiometric correction to homogenize large-scale intensity between two SSS images acquired from adjacent survey lines, denoted as the \textit{fixed image} $I_f$ and the \textit{moving image} $I_m$. This is followed by TPS-based global warping to establish a coarse geometric alignment, providing a suitable initialization for subsequent local refinement.

\subsubsection{Swath-Wise Radiometric Correction}

To homogenize the intensity distribution across the swath, we adopt a two-step approach combining column-wise radiometric correction and local contrast enhancement.

\paragraph{Column-wise amplitude correction}

Given a grayscale SSS image $I \in \mathbb{R}^{H \times W}$, the mean intensity of each column is computed as:
\begin{equation}
    {{\alpha }_{j}}=\frac{{\bar{\mu }}}{{{\mu }_{j}}},\bar{\mu }=\frac{1}{W}\sum\limits_{j=1}^{W}{{{\mu }_{j}}},{{\mu }_{j}}=\frac{1}{H}\sum\limits_{i=1}^{H}{{{I}_{i,j}}}
\end{equation}
where $I_{i,j}$ denotes the pixel intensity at row $i$ and column $j$, $H$ and $W$ are the image height and width, respectively. A scaling matrix $M_{i,j} = \alpha_j$ is then applied to yield the radiometrically corrected image $I^{\text{corr}}$ with uniform intensity:
\begin{equation}
    I^{\text{corr}} = I \odot M
\end{equation}
where $\odot$ denotes element-wise multiplication.

\paragraph{Residual variation suppression and local enhancement}

Weighted local contrast enhancement is applied after range-decay correction:
\begin{equation}
    \overset{\lower0.5em\hbox{$\smash{\scriptscriptstyle\smile}$}}{I}={{I}^{\text{corr}}}+\lambda \left( {{I}^{\text{corr}}}-{{\mathcal{G}}_{\sigma }}({{I}^{\text{corr}}}) \right)
\end{equation}
where $\mathcal{G}{\sigma}(\cdot)$ is a Gaussian smoothing operation with standard deviation $\sigma=15$, and $\lambda=0.5$ controls enhancement strength. This step attenuates large-scale radiometric variations while preserving and amplifying fine-scale textures, producing the corrected images ${{\overset{\lower0.5em\hbox{$\smash{\scriptscriptstyle\smile}$}}{I}}_{f}}$ and ${{\overset{\lower0.5em\hbox{$\smash{\scriptscriptstyle\smile}$}}{I}}_{m}}$ for subsequent alignment.

\subsubsection{Sparse Control Point-Based TPS Alignment}

After radiometric correction, a coarse global alignment is performed using a TPS transformation estimated from sparse, manually selected control points within overlapping regions of the corrected fixed and moving images.

Let ${(\mathbf{p}_i, \mathbf{q}i)}{i=1}^N$ denote the matched control point pairs in the corrected fixed image ${{\overset{\lower0.5em\hbox{$\smash{\scriptscriptstyle\smile}$}}{I}}_{f}}$ and moving image ${{\overset{\lower0.5em\hbox{$\smash{\scriptscriptstyle\smile}$}}{I}}_{m}}$, respectively. The TPS transformation $\Phi_{\text{TPS}}: \mathbb{R}^2 \to \mathbb{R}^2$ is defined as:
\begin{equation}
   \begin{array}{l}
   {{\Phi }_{\text{TPS}}}(\mathbf{x})=\mathbf{Ax}+\mathbf{b}+\sum\limits_{i=1}^{N}{{{\mathbf{w}}_{i}}}U(\left\| \mathbf{x}-{{\mathbf{q}}_{i}} \right\|),U(r)={{r}^{2}}\log r \\ 
  s.t.\quad {{\Phi }_{\text{TPS}}}({{\mathbf{q}}_{i}})={{\mathbf{p}}_{i}},\mathbf{x}\in {{\mathbb{R}}^{2}} \\ 
    \end{array}
\end{equation}
where $\mathbf{A} \in \mathbb{R}^{2\times2}$ and $\mathbf{b} \in \mathbb{R}^2$ are affine coefficients, and $\mathbf{w}i \in \mathbb{R}^2$ are non-linear kernel weights. The parameters $(\mathbf{A},\mathbf{b},\{\mathbf{w}_i\})$ are solved so that all control points are exactly matched, and the radial basis kernel $U(\cdot)$ enforces a smooth spatial interpolation across the image domain.

Applying $\Phi_{\mathrm{TPS}}$ to ${{\overset{\lower0.5em\hbox{$\smash{\scriptscriptstyle\smile}$}}{I}}_{m}}$ yields the globally warped moving image:
\begin{equation}
    \overset{\lower0.5em\hbox{$\smash{\scriptscriptstyle\smile}$}}{I}_{m}^\text{warp}(\mathbf{x})={{{\overset{\lower0.5em\hbox{$\smash{\scriptscriptstyle\smile}$}}{I}}}_{m}}\left( {{\Phi }_{\text{TPS}}}(\mathbf{x}) \right)
\end{equation}
where $\overset{\lower0.5em\hbox{$\smash{\scriptscriptstyle\smile}$}}{I}_{m}^\text{warp}$ comprises an overlapping region $\overset{\lower0.5em\hbox{$\smash{\scriptscriptstyle\smile}$}}{I}_{m_{ov}}^\text{warp}$—hereafter referred to as the moved image—and a non-overlapping region $\overset{\lower0.5em\hbox{$\smash{\scriptscriptstyle\smile}$}}{I}_{m_{non}}^\text{warp}$. Since the non-overlapping region has no counterpart in the fixed image, only the overlapping part is passed to the subsequent local non-rigid registration stage.

\subsection{Local Patch Generation via Superpixel Segmentation}

\begin{figure*}[htbp]
	\centering
	\includegraphics[width=1\linewidth]{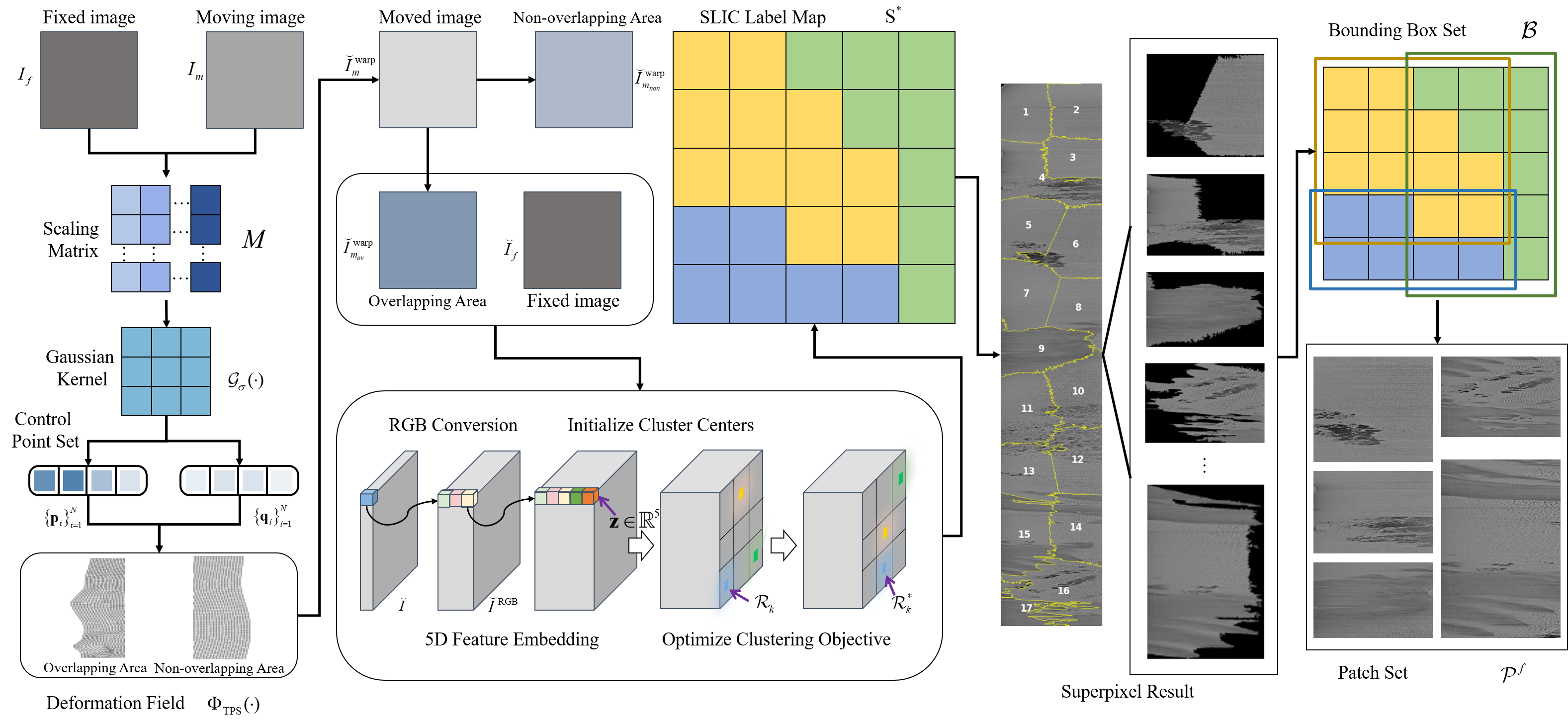}
 	\caption{Initial global non-rigid alignment and local patch generation pipeline. Radiometric correction mitigates range-dependent intensity inhomogeneity through column-wise amplitude scaling and local contrast enhancement. Sparse control points in overlapping regions guide TPS estimation to produce a coarse global deformation field, aligning the moving image to the fixed image. Subsequently, superpixel segmentation on the fixed image generates structurally consistent regions, which are enclosed by bounding boxes to define regular local patches. Corresponding patch pairs are extracted from the fixed and globally warped moving images, forming the basis for localized non-rigid refinement.}
	\label{fig01}
\end{figure*}

After the initial global alignment via TPS deformation, residual misalignments remain in overlapping regions that require localized refinement. To achieve this, we employ superpixel segmentation to adaptively partition the global image into structurally consistent regions for local registration (Fig.~\ref{fig01}).

\subsubsection{Superpixel Segmentation}

Starting from the corrected fixed image ${{{\overset{\lower0.5em\hbox{$\smash{\scriptscriptstyle\smile}$}}{I}}}_{f}}$, which is originally grayscale, we convert it into a pseudo-RGB image $\overset{\lower0.5em\hbox{$\smash{\scriptscriptstyle\smile}$}}{I}_{f}^{\text{RGB}}\in {{\mathbb{R}}^{H\times W\times 3}}$ by replicating the intensity values across the three channels,  a conversion that is crucial for effective superpixel segmentation. Each pixel ${{\mathbf{x}}_{i}}$ in $\overset{\lower0.5em\hbox{$\smash{\scriptscriptstyle\smile}$}}{I}_{f}^{\text{RGB}}$ is represented as a five-dimensional feature vector ${{\mathbf{z}}_{i}}=[\overset{\lower0.5em\hbox{$\smash{\scriptscriptstyle\smile}$}}{I}_{f}^{\text{RGB}}\,({{\mathbf{x}}_{i}}),\ {{\mathbf{x}}_{i}}]\in {{\mathbb{R}}^{5}}$, concatenating its color and spatial coordinates. Based on these features, we perform iterative clustering that minimizes a combined color and spatial distance objective:
\begin{equation}
    \mathcal{L}(S)=\underset{k=1}{\overset{K}{\mathop \sum }}\,\underset{{{\mathbf{x}}_{i}}\in {{\mathcal{R}}_{k}}}{\mathop \sum }\,\parallel \overset{\lower0.5em\hbox{$\smash{\scriptscriptstyle\smile}$}}{I}_{f}^\text{RGB}({{\mathbf{x}}_{i}})-{{\mu }_{k}}\parallel _{2}^{2}+{{\lambda }_{s}}\parallel {{\mathbf{x}}_{i}}-{{\mathbf{c}}_{k}}\parallel _{2}^{2}
\end{equation}
where $\mathcal{R}_k$ denotes the $k$-th superpixel region with mean color $\mu_k$ and centroid $\mathbf{c}_k$, and $\lambda_s$ is a compactness factor balancing color similarity and spatial compactness.

By optimizing $\mathcal{L}$, we obtain the label map ${{S}^{*}}\in {{\mathbb{R}}^{H\times W}}$ that partitions ${{{\overset{\lower0.5em\hbox{$\smash{\scriptscriptstyle\smile}$}}{I}}}_{f}}$ into structurally consistent local regions $\left\{ \mathcal{R}_{k}^{*},k=1,\ldots ,K \right\}$. This segmentation result serves as the foundation for defining bounded patches in the fixed image, which are then correspondingly mapped onto the moved image $\overset{\lower0.5em\hbox{$\smash{\scriptscriptstyle\smile}$}}{I}_{{{m}_{ov}}}^\text{warp}$ to establish spatially consistent patch pairs for registration.

\subsubsection{Bounding Box Computation}

To ensure stable local registration, each irregular superpixel region is converted into a regular rectangular patch. For each superpixel region $\mathcal{R}_{k}^{*}$, the smallest axis-aligned bounding box enclosing the region is computed as:
\begin{equation}
{{\mathcal{B}}_{k}}=[r_{\min }^{k},\ r_{\max }^{k},\ c_{\min }^{k},c_{\max }^{k}]
\end{equation}
where $r_{\min}^{k}$ and $r_{\max}^{k}$  denote the minimum and maximum row indices, and $c_{\min}^{k}$ and $c_{\max}^{k}$ denote the minimum and maximum column indices of region $\mathcal{R}_{k}^{*}$, respectively. The full set of bounding boxes is defined as: $\mathcal{B}=\left\{ {{\mathcal{B}}_{k}},k=1,\ldots ,K \right\}$.

Due to the proximity of adjacent superpixels, these rectangular patches partially overlap. This intentional overlap allows shared content near region boundaries to appear in multiple patch pairs, providing diverse local observations that benefit the subsequent fusion of deformation fields.

\subsubsection{Aligned Patch Pairs Extraction}

Based on ${ \mathcal{B} }$, we extract corresponding local patches from both ${{{\overset{\lower0.5em\hbox{$\smash{\scriptscriptstyle\smile}$}}{I}}}_{f}}$ and $\overset{\lower0.5em\hbox{$\smash{\scriptscriptstyle\smile}$}}{I}_{{{m}_{ov}}}^\text{warp}$. For each bounding box ${{\mathcal{B}}_{k}}$, the patches are defined as:
\begin{equation}
   \begin{array}{l}
   P_{k}^{f}={{{\overset{\lower0.5em\hbox{$\smash{\scriptscriptstyle\smile}$}}{I}}}_{f}}[r_{\min }^{k}:\ r_{\max }^{k},\ c_{\min }^{k}:c_{\max }^{k}] \\ 
  P_{k}^{{{m}_{ov}}}=\overset{\lower0.5em\hbox{$\smash{\scriptscriptstyle\smile}$}}{I}_{{{m}_{ov}}}^\text{warp}[r_{\min }^{k}:\ r_{\max }^{k},\ c_{\min }^{k}:c_{\max }^{k}] \\ 
\end{array}
\end{equation}

The sets of extracted patches from the fixed and moved images are then collected as:

\begin{equation}
    \begin{array}{l}
   {{\mathcal{P}}^{f}}=\{P_{k}^{f}\mid k=1,\ldots ,K\} \\ 
 {{\mathcal{P}}^{{{m}_{ov}}}}=\{P_{k}^{{{m}_{ov}}}\mid k=1,\ldots ,K\} \\ 
    \end{array}
\end{equation}
where $P_{k}^{f}$ denotes the $k$-th patch extracted from the fixed image, and $P_{k}^{{{m}_{ov}}}$ denotes the corresponding patch from the moved image (warped overlapping). Each patch pair $(P_{k}^{f},P_{k}^{{{m}_{ov}}})$ corresponds to a spatially aligned local region and is independently used for subsequent fine-grained local registration.

\subsection{Patch-Wise Non-Rigid Registration}

While the preceding global registration achieves coarse alignment, it cannot fully resolve fine-scale distortions caused by complex local deformations and imaging artifacts in SSS image. To address this, we adopt a patch-wise non-rigid registration pipeline: first applying artifact removal to enhance patch quality, then estimating dense local deformation fields for alignment (Fig.~\ref{fig02}).

\subsubsection{Preprocessing for Local Registration}

To improve registration accuracy, stripe artifacts from channel imbalance and TVG compensation are removed through stripe suppression and structural enhancement, preserving features essential for alignment.

\paragraph{Stripe Artifact Suppression}

Unlike random noise, stripe artifacts in SSS images exhibit long-range coherence, making local filters ineffective. So we apply non-local means filtering exploiting patch self-similarity to suppress these artifacts.

Given an input patch ${{P}_{k}}\in {{\mathbb{R}}^{{H}'\times {W}'}}$, the denoised output $\hat{P}_k$ is computed by weighted averaging over a local search window:
\begin{equation}
    \begin{array}{l}
  {{{\hat{P}}}_{k}}(\mathbf{x})=\frac{1}{Z(\mathbf{x})}\sum\limits_{\mathbf{y}\in \Omega (\mathbf{x})}{w(\mathbf{x,y})\cdot {{P}_{k}}(\mathbf{y})} \\ 
 w(\mathbf{x,y})=\exp \left( -\frac{\|{{P}_{k}}({{\mathcal{N}}_{\mathbf{x}}})-{{P}_{k}}({{\mathcal{N}}_{\mathbf{y}}})\|_{2}^{2}}{{{h}^{2}}} \right) \\ 
  Z(\mathbf{x})=\sum\limits_{\mathbf{y}\in \Omega (\mathbf{x})}{w(\mathbf{x,y})} \\ 
    \end{array}
\end{equation}
where $\mathbf{x}, \mathbf{y} \in \mathbb{R}^2$ denote pixel coordinates, $\mathcal{N}_{\mathbf{x}}$ denotes a $7 \times 7$ patch centered at $\mathbf{x}$, and $\mathcal{N}_{\mathbf{y}}$ is defined similarly. $\Omega(\mathbf{x})$ denotes a $21 \times 21$ square search window centered at $\mathbf{x}$. $w(\mathbf{x}, \mathbf{y})$ is the similarity weight between patches centered at $\mathbf{x}$ and $\mathbf{y}$, $Z(\mathbf{x})$ is the normalization term, and $h=10$ balances denoising strength and detail preservation, thereby preserving critical features such as seafloor texture, which are vital for robust registration.

\paragraph{Laplacian-Based Structural Enhancement}

Non-local filtering, while suppressing stripe noise, tends to blur high-frequency details and reduce edge contrast. To restore these vital features, a Laplacian-based sharpening filter is applied to the denoised patch $\hat{P}_k$ to obtain the enhanced patch $\tilde{P}_k$:
\begin{equation}
   \begin{array}{l}
   {{{\tilde{P}}}_{k}}={{{\hat{P}}}_{k}}*{{K}_{\text{sharp}}}={{{\hat{P}}}_{k}}*(\delta +L)={{{\hat{P}}}_{k}}+{{{\hat{P}}}_{k}}*L \\ 
  \delta =\left[ \begin{matrix}
   0 & 0 & 0  \\
   0 & 1 & 0  \\
   0 & 0 & 0  \\
\end{matrix} \right],L=\left[ \begin{matrix}
   0 & -1 & 0  \\
   -1 & 4 & -1  \\
   0 & -1 & 0  \\
\end{matrix} \right] \\ 
\end{array}
\end{equation}
here, $\delta$ is the identity kernel, $L$ is the Laplacian operator, $*$ denotes convolution, and $K_{\text{sharp}}$ is the composite sharpening kernel; convolution with $K_{\text{sharp}}$ enhances local intensity transitions and edge saliency.

For each patch pair $(P_{k}^{f},P_{k}^{{m}_{ov}})$ derived from superpixel segmentation, Sequential Stripe Suppression and Structural Enhancement are applied to generate enhanced representations $(\tilde{P}_{k}^{f},\tilde{P}_{k}^{{m}_{ov}})$, which are then fed into the local registration network.

\subsubsection{Local Non-Rigid Registration}

\begin{figure*}[htbp]
	\centering
	\includegraphics[width=1\linewidth]{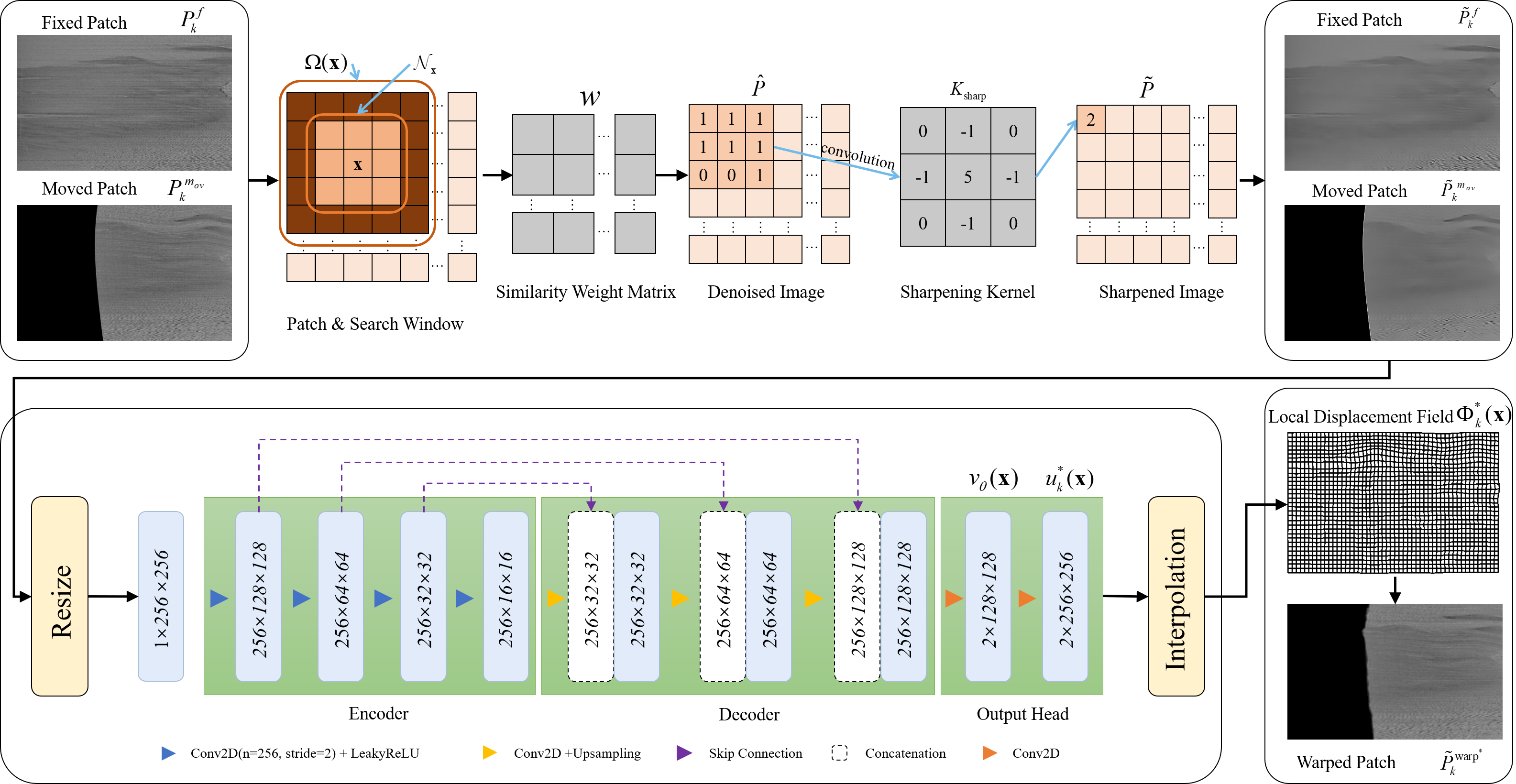}
	\caption{Patch-wise non-rigid registration workflow. Each superpixel-derived patch pair undergoes preprocessing to suppress stripe artifacts via non-local means filtering and enhance structural details using a Laplacian sharpening filter. The preprocessed patches, resized to fixed dimensions, are input to an unsupervised SynthMorph network that predicts dense local displacement fields. These displacement fields warp the moved patches to align with fixed patches. The resulting local deformation fields capture fine-scale geometric variations essential for precise alignment.}
	\label{fig02}
\end{figure*}	

The preprocessed patch pairs are then input to an unsupervised deep network that performs dense local registration, producing deformation fields for each patch to capture fine-scale geometric variations.

\paragraph{Network Architecture for Local Registration}

Given a processed patch pair $(\tilde{P}_{k}^{f},\tilde{P}_{k}^{{m}_{ov}})$, we first resize both patches to a fixed size of $256\times 256$ to conform with the network input requirements. The resized moved patch is then fed into the SynthMorph network \cite{A37}, which predicts a dense 2D displacement field ${{u}_{k}}(\mathbf{x})\in {{\mathbb{R}}^{2}}$ that registers the moved patch to the fixed patch.

Built upon a U-Net architecture, the network encodes the input $\tilde{P}_{k}^{{m}_{ov}}$ through four stages of \texttt{stride-2 Convolution} layers, each with 256 filters and \texttt{LeakyReLU} activation, progressively downsampling spatial resolution while capturing multi-scale features. The decoder symmetrically \texttt{upsamples} through three stages with \texttt{skip connections} from corresponding encoder layers. The output head first predicts a stationary velocity field $v_\theta(\mathbf{x})$ at half input resolution, which is converted via a final convolution layer into the displacement field ${{u}_{k}}(\mathbf{x})$. After inference, the predicted displacement field is interpolated back to the original patch size $({H}',{W}')$ to maintain spatial consistency. Fig. \ref{fig02} illustrates the detailed network architecture.

\paragraph{Registration Objective and Deformation Estimation}

With the displacement field ${{u}_{k}}(\mathbf{x})$, the warped moved patch $\tilde{P}_{k}^\text{warp}$ is computed by applying bilinear interpolation:
\begin{equation}
    \tilde{P}_{k}^\text{warp}(\mathbf{x})=\mathcal{I}(\tilde{P}_{k}^{{{m}_{ov}}},{{\Phi }_{k}}(\mathbf{x})),{{\Phi }_{k}}(\mathbf{x})=\mathbf{x}+{{u}_{k}}(\mathbf{x})
\end{equation}
where ${{\Phi }_{k}}(\mathbf{x})$ denotes the dense deformation field and $\mathcal{I}$ is the bilinear interpolation operator.

To estimate an accurate and plausible deformation, the loss function combines a mean squared error between $\tilde{P}_{k}^\text{warp}$ and the fixed patch $\tilde{P}_{k}^{f}$, and a spatial regularization term that encourages smooth and coherent deformation. The final predicted deformation field $\Phi _{k}^{*}$ is obtained by minimizing the following objective:
\begin{equation}
    \Phi _{k}^{*}=\arg {{\min }_{\Phi }}\ \mathcal{D}(\tilde{P}_{k}^{f},\tilde{P}_{k}^\text{warp})+\mathcal{R}(\Phi )
\end{equation}
where $\mathcal{D}(\cdot )$ denotes the mean squared error and $\mathcal{R}(\cdot )$ penalizes non-smooth warping fields. After optimization, the final warped patch $\tilde{P}{{_{k}^{\text{warp}}}^{*}}$ is computed by applying the optimal deformation field $\Phi _{k}^{*}$ to the moved patch using bilinear interpolation:
\begin{equation}
  \tilde{P}{{_{k}^{\text{warp}}}^{*}}(\mathbf{x})=\mathcal{I}(\tilde{P}_{k}^{{{m}_{ov}}},\Phi _{k}^{*}(\mathbf{x})), u_{k}^{*}(\mathbf{x})=\Phi _{k}^{*}(\mathbf{x})-\mathbf{x} 
\end{equation}
here, $u_k^*(\mathbf{x})$ denotes the final predicted local displacement filed at location $\mathbf{x}$, which will be used for the subsequent fusion of local deformation fields into a global registration map. In contrast, the warped patch $\tilde{P}{{_{k}^{\text{warp}}}^{*}}$ serves only for local visualization and validation, as directly fusing warped image patches would result in visible seams.

Importantly, we leverage a pretrained SynthMorph model trained in an unsupervised fashion on large-scale multimodal datasets. Its strong generalization enables direct application to sonar imagery without additional fine-tuning, effectively supporting accurate local non-rigid registration within our pipeline.

\subsection{Deformation Field Aggregation and Global Warping}

Building upon the fine-grained local deformation fields obtained from patch-wise registration, we next integrate these into a coherent global deformation. This involves aggregating overlapping local fields and fusing them with an initial TPS-based global transformation, producing the final deformation field for subsequent image warping and mosaic composition.

\subsubsection{Aggregation of Local Displacements}

Given each registered patch pair $(\tilde{P}_{k}^{f},\tilde{P}_{k}^{{{m}_{ov}}})$, the local displacement field is defined as:
\begin{equation}
    u_k^*(\mathbf{x}) = \left( u_k^x(\mathbf{x}),\ u_k^y(\mathbf{x}) \right)
\end{equation}
where $u_{k}^{x}(\mathbf{x})$ and $u_k^y(\mathbf{x})$ denote the horizontal and vertical displacement components over the patch domain ${{\Gamma }_{k}}\subset {{\mathbb{R}}^{2}}$. Due to the superpixel-based segmentation strategy, adjacent patches partially overlap.

To merge all local estimates into a dense displacement field over the full image domain $\Gamma$, we adopt a Gaussian-weighted averaging:
\begin{equation}
   \begin{array}{l}
   {{u}^{x}}(x,y)=\frac{\sum\limits_{k}{{{G}_{k}}}(x,y)\cdot u_{k}^{x}(x,y)}{\sum\limits_{k}{{{G}_{k}}}(x,y)+\epsilon } \\ 
  {{u}^{y}}(x,y)=\frac{\sum\limits_{k}{{{G}_{k}}}(x,y)\cdot u_{k}^{y}(x,y)}{\sum\limits_{k}{{{G}_{k}}}(x,y)+\epsilon }
    \end{array}
\end{equation}
where $G_k(x,y)$ is a 2D Gaussian weight map centered at the $k$-th patch, emphasizing central pixels and attenuating boundaries to ensure smooth blending. $u^x(x,y)$ and $u^y(x,y)$ are a pair of displacement maps that jointly represent the full-resolution, locally refined deformation over the overlapping regions. A small $\epsilon$ prevents division-by-zero.

\subsubsection{Fusion with TPS-Based Global Deformation}

The coarse-scale TPS deformation $\Phi_{\text{TPS}}(x,y) = \left( \Phi_{\text{TPS}}^{x}(x,y),\ \Phi_{\text{TPS}}^{y}(x,y) \right)$ provides large-scale alignment, while the fused local displacement field $\left(u^{x}(x,y), u^{y}(x,y)\right)$ refines small-scale distortions. Since the local displacements are estimated within the overlapping region of the TPS-warped moving image $\overset{\lower0.5em\hbox{$\smash{\scriptscriptstyle\smile}$}}{I}_{m}^\text{warp}$ and the fixed image, fusion is performed only within the overlapping region $\Gamma_{{ov}}$, whereas the non-overlapping region $\Gamma_{{non}}$ retains the original TPS deformation.

Therefore, the final composite deformation field  ${{\Phi }_{\text{final}}}$ is obtained as:
\begin{small}
\begin{equation}
{{\Phi }_{\text{final}}}(x,y)=\left\{\begin{array}{l}
   \left( \Phi _{\text{TPS}}^{x}(x,y)+{{u}^{x}}(x,y),\Phi _{\text{TPS}}^{y}(x,y)+{{u}^{y}}(x,y) \right),(x,y)\in {{\Gamma }_{{ov}}} \\ 
  \left(\Phi _{\text{TPS}}^{x}(x,y),\Phi _{\text{TPS}}^{y}(x,y) \right),(x,y)\in {{\Gamma }_{{non}}} \\ 
\end{array} \right.
\end{equation}
\end{small}

This selective fusion preserves the stability of coarse global alignment in non-overlapping areas, while fine-grained local corrections are integrated into the overlapping regions, producing a smooth and spatially coherent deformation field across the entire sonar image domain.

\subsubsection{Final Image Warping and Mosaic Composition}

With the final deformation field $\Phi_{\text{final}}$ obtained, the moving image $I_m$ is spatially transformed via bilinear interpolation:
\begin{equation}
    I_{m}^{\text{warp}^*}(x,y)=\left\{ \begin{array}{l}
   I_{{{m}_{{ov}}}}^{\text{warp}^*}(x,y),(x,y)\in {{\Gamma }_{{ov}}} \\ 
  I_{{{m}_{{non}}}}^{\text{warp}^*}(x,y),(x,y)\in {{\Gamma }_{{non}}} \\ 
\end{array} \right.
\end{equation}

The registered moving image $ I_{m}^{\text{warp}^*}$ is then integrated with the fixed image $I_f$ to generate the final mosaic ${{I}_{\text{mos}}}$. In the overlapping region $\Gamma_{\text{ov}}$, pixel intensities from the two images are fused using a pixel-wise maximum operation to preserve salient features:
\begin{equation}
    {{I}_{\text{mos}}}(x,y)=\left\{ \begin{array}{l}
   \max ({{I}_{f}}(x,y), I_{m}^{\text{warp}^*}(x,y)),(x,y)\in {{\Gamma }_{{ov}}} \\ 
  I_{{{m}_{{non}}}}^{\text{warp}^*}(x,y),(x,y)\in {{\Gamma }_{{non}}} \\ 
\end{array} \right.
\end{equation}

For the non-overlapping region $\Gamma_{{non}}$, the mosaic directly inherits pixel values from the available image, ensuring spatial continuity without introducing artificial boundaries, thus maintaining high-fidelity structural details in overlapping areas and smooth geometric extension into unmapped regions.

\section{Experiment}

\subsection{Data Description}

\subsubsection{Data Acquisition}

 The experimental data employed in this study originates from sector N07 of the dataset introduced in \cite{A46}. The sonar imagery was collected using a Klein 3000H SSS system operating at a center frequency of \SI{500}{\kilo\hertz}. During data acquisition, the sensor range was manually configured according to the local bathymetric conditions, ranging from \SI{50}{\meter} and \SI{100}{\meter} across different survey areas. The platform altitude was maintained at approximately \SI{10}{\percent} of the set range, yielding an average slant-range resolution of \SI{8.5}{\centi\meter}.

\subsubsection{Data Preprocessing}

To prepare the raw .xtf side-scan sonar data for subsequent registration processing, we extract intensity values per ping and apply a logarithmic transformation ${I}'={{\log }_{10}}(I+\epsilon )$ with $\epsilon =1e-6$ to enhance dynamic range and contrast while avoiding undefined values. The transformed intensities are then normalized to the $[0,1]$ range to ensure consistency and improve robustness in the following registration steps.

\begin{figure*}[htbp]
	\centering
	\includegraphics[width=1\linewidth]{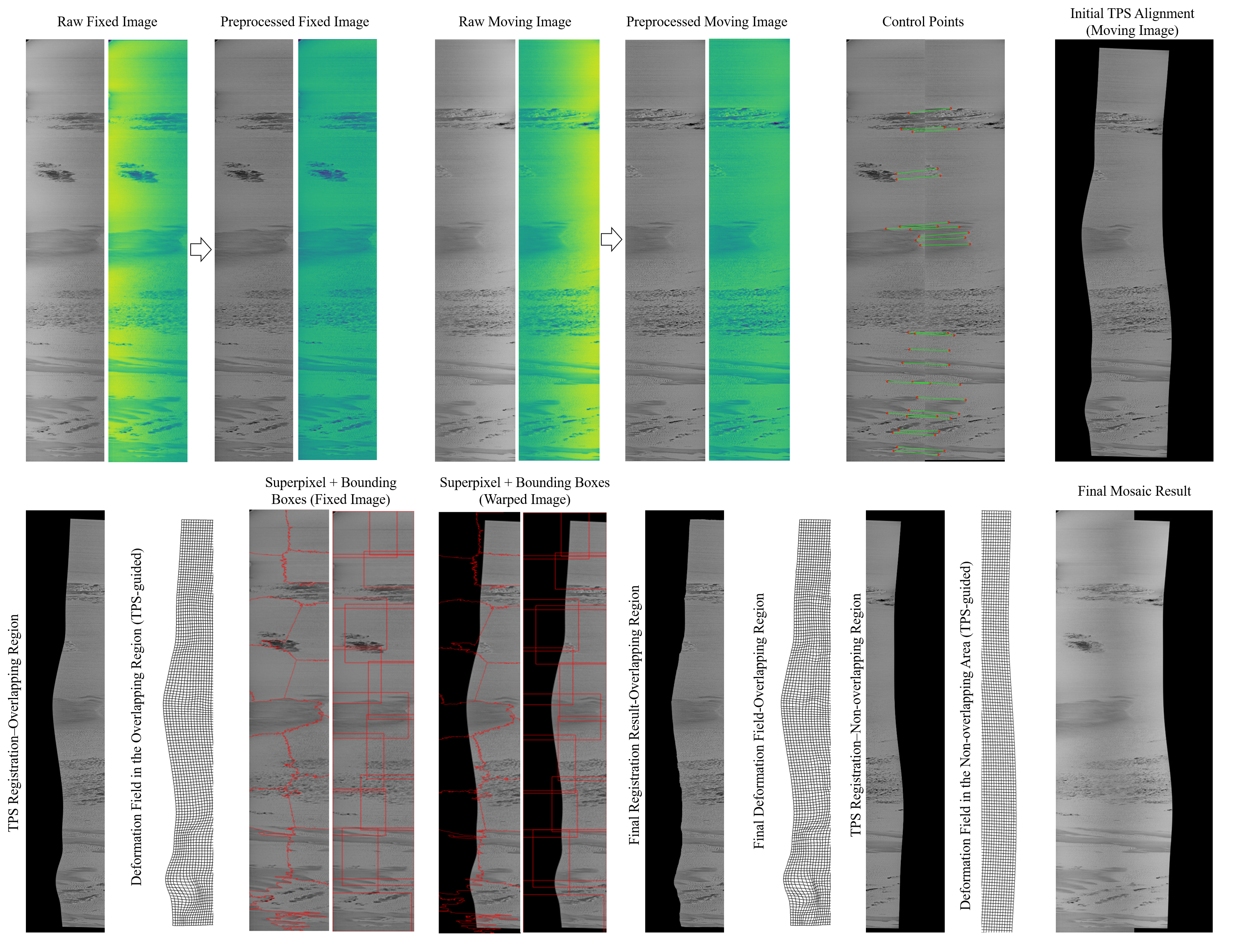}
	\caption{Visualization of the global registration process. Top row: Left — original and preprocessed fixed/moving images in grayscale and pseudo-color formats (where higher intensities appear yellow-green, lower intensities appear blue-green). Right — manually selected control point pairs and the resulting initial TPS alignment on the moving image. Bottom row: Left — separation of overlapping and non-overlapping regions after initial TPS registration. Overlapping results and deformation fields are shown first, followed by superpixel segmentation and bounding box partitioning on both the fixed and warped images. Middle — refined global registration result (overlapping region only) and its deformation field. Right — extrapolated non-overlapping results from TPS and final mosaic output. }
	\label{fig0}
\end{figure*}	

\begin{figure}[htbp]
	\centering
	\includegraphics[width=1\linewidth]{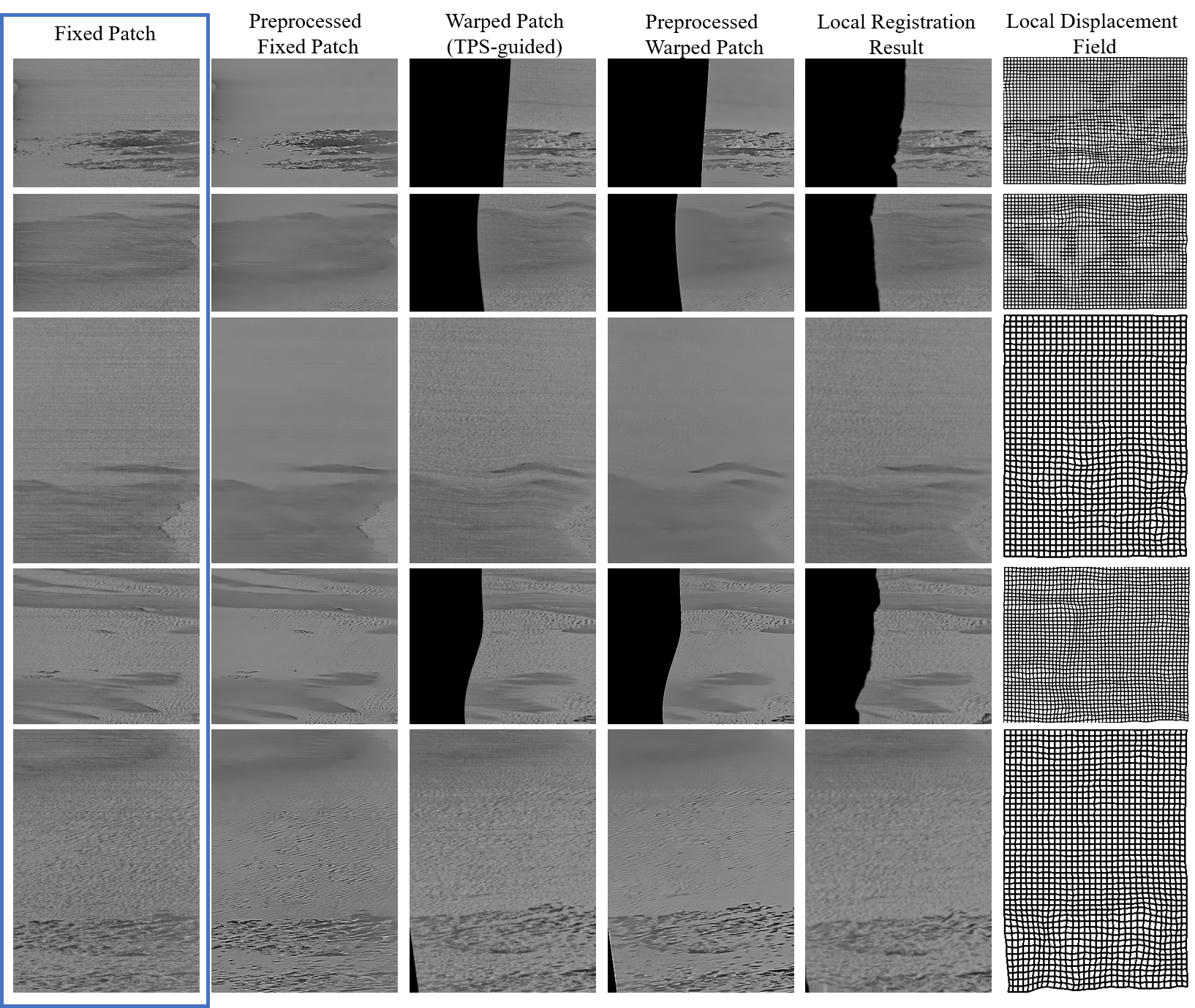}
	\caption{Local visualization of the patch-wise non-rigid registration process. This figure presents local registration results across 5 selected patch regions (rows). Each row includes the following columns: (1) Fixed patch (reference); (2) Preprocessed fixed patch; (3) Warped patch (extracted from globally warped image via bounding box); (4) Preprocessed warped patch; (5) Locally registered patch using SynthMorph;(6) Corresponding local displacement field (visualized as a dense displacement grid).}
	\label{fig1}
\end{figure}	

\begin{figure}[htbp]
	\centering
	\includegraphics[width=1\linewidth]{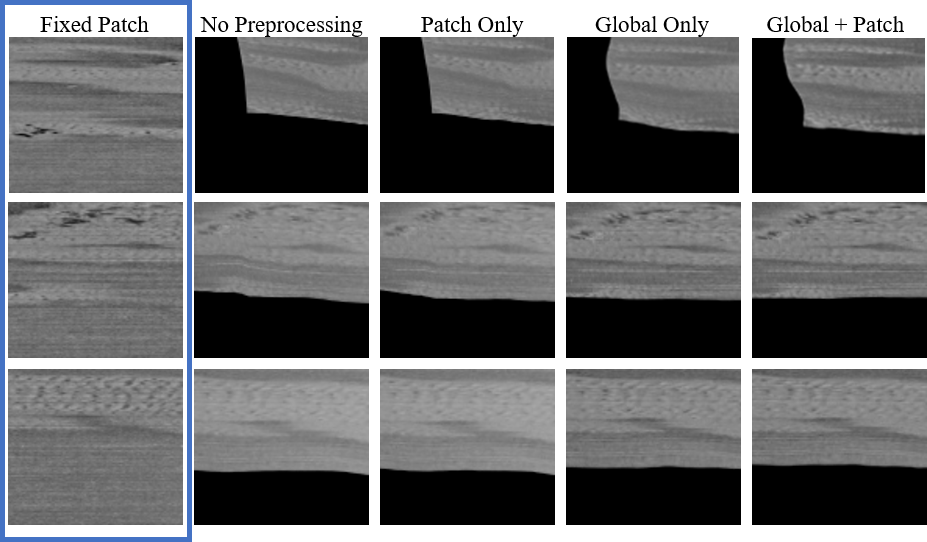}
	\caption{Visual comparison of local registration results under different preprocessing strategies. Each row corresponds to a different patch region selected from overlapping areas. Column 1: Reference fixed patch. Column 2: Registration result without any preprocessing.  Column 3: Registration result with patch-level preprocessing only.  Column 4: Registration result with global-level preprocessing only.  Column 5: Registration result with both global and patch-level preprocessing.}
	\label{fig2}
\end{figure}

\begin{table*}
	\caption{Ablation Results under Different Preprocessing Strategies and Superpixel Numbers.}
    \vspace{1mm}
    \small \textbf{Note:}
    ${\bar{e}}$ denotes the average control point error (pixels) measuring geometric alignment accuracy; ${R}_{{ov}}$ is the overlap ratio (\%) reflecting overlapping area; $SSIM$ and $MI$ evaluate structural and statistical similarity; $RMSE$ quantifies pixel-wise intensity differences; ${R}_{\text{fold}}$ indicates the percentage of folding regions in the deformation field; and ${\sigma}_{\text{disp}}$ measures the standard deviation of displacement magnitudes.
    
	\label{tab1}
	\centering
	\footnotesize
	\setlength{\tabcolsep}{19pt}
	\renewcommand{\arraystretch}{1.2}
	\begin{tabular}{lccccccc}
		\toprule
		\textbf{Method} & \textbf{${\bar{e}}$(px)} & \textbf{${{R}_{{ov}}}$} & \textbf{$SSIM$} & \textbf{$MI$} & \textbf{$RMSE$} & \textbf{${{R}_{\text{fold}}}$} & \textbf{${{\sigma }_{\text{disp}}}$} \\
        \midrule
		\textbf{TPS-10} & & & & & & & \\
 No Preprocessing  & 9.37  & 39.09\% & 0.62 & 0.74 & 16.17 & 35.73\% & 125.93 \\
 Patch Only        & 8.90  & 39.16\% & 0.62 & 0.74 & 16.16 & 35.65\% & 125.97 \\
 Global Only       & 8.35  & 52.23\% & 0.68 & 0.75 & 14.96 & 34.91\% & 126.08 \\
 Global + Patch    & 8.30  & 52.25\% & 0.68 & 0.75 & 14.95 & \textbf{34.66\%} & 126.11 \\
        \midrule
        \textbf{TPS-20} & & & & & & & \\
 No Preprocessing  & 8.50  & 39.65\% & 0.63 & 0.74 & 16.09 & 37.87\% & 125.94 \\
 Patch Only        & 8.23  & 39.80\% & 0.63 & 0.74 & 16.03 & 37.74\% & 125.94 \\
 Global Only       & 7.64  & 52.79\% & 0.69 & 0.75 & 14.88 & 37.34\% & 126.20 \\
 Global + Patch    & 6.87  & 52.92\% & 0.69 & 0.75 & \textbf{14.86} & 36.95\% & 126.23 \\
        \midrule
        \textbf{TPS-30} & & & & & & & \\
 No Preprocessing  & 7.73  & 39.54\% & 0.63 & 0.75 & 16.07 & 38.79\% & 125.97 \\
 Patch Only        & 7.17  & 39.74\% & 0.63 & 0.75 & 16.00 & 38.70\% & 125.96 \\
 Global Only       & 6.15  & 53.28\% & 0.69 & 0.75 & 14.89 & 37.79\% & 126.17 \\
 Global + Patch    & \textbf{5.12}  & \textbf{53.41\%} & 0.69 & 0.75 & 14.87 & 37.64\% & 126.20 \\
        \midrule
        \textbf{TPS-40} & & & & & & & \\
 No Preprocessing  & 7.88  & 39.37\% & 0.63 & 0.74 & 16.14 & 37.69\% & 125.94 \\
 Patch Only        & 6.35  & 39.72\% & 0.63 & 0.74 & 16.06 & 37.45\% & 125.96 \\
 Global Only       & 6.30  & 52.93\% & 0.69 & 0.75 & 14.88 & 37.29\% & 126.19 \\
 Global + Patch    & 6.19  & 52.95\% & 0.69 & 0.75 & 14.85 & 37.10\% & 126.23 \\
        \bottomrule
	\end{tabular}
\end{table*}

\subsection{Evaluation Metrics}

\subsubsection{Control-point error}

A set of $N$ manually annotated corresponding control points, independent from the control points used by TPS, is selected for each image pair ($N\approx30$). Denote the fixed/moving coordinates as $\{(x_i^{\mathrm{fix}},y_i^{\mathrm{fix}}),(x_i^{\mathrm{mov}},y_i^{\mathrm{mov}})\}_{i=1}^N$. The final deformation field $\Phi_{\mathrm{final}}$ warps moving points to
$(\hat{x}_i^{\mathrm{fix}},\hat{y}_i^{\mathrm{fix}})=\Phi_{\mathrm{final}}(x_i^{\mathrm{mov}},y_i^{\mathrm{mov}})$. 
Per-point Euclidean registration error is
\begin{equation}
    e_i=\sqrt{(x_i^{\mathrm{fix}}-\hat{x}_i^{\mathrm{fix}})^2+(y_i^{\mathrm{fix}}-\hat{y}_i^{\mathrm{fix}})^2}
\end{equation}

We report the mean registration error: $\bar{e}=\frac{1}{N}\sum\limits_{i=1}^{N}{{{e}_{i}}}$. The independence of these control points from TPS anchors ensures this metric evaluates the full pipeline without bias.

\subsubsection{Overlap-Based Consistency Metrics}

To evaluate spatial and structural consistency after non-rigid registration, we define four complementary metrics, all computed within the overlapping region $\Gamma _{ov}$ between the fixed image $I_f$ and the final warped moving image $I_{m_{ov}}^{\text{warp}^*}$.

\paragraph{Overlap Ratio} 

The overlap mask is defined by
\begin{equation}
M_{ov}(x,y) =
\begin{cases}
1, & \big| I_f(x,y) - I_{m_{ov}}^{\text{warp}^*}(x,y) \big| \le \delta \\
0, & \text{otherwise}
\end{cases}
\end{equation}
where $\delta$ is an intensity tolerance. The overlap ratio is
\begin{equation}
\begin{array}{l}
   {{R}_{{ov}}}=\frac{\sum\limits_{(x,y)\in {{R}_{\text{valid}}}}{{{M}_{\text{ov}}}}(x,y)}{|{{R}_{\text{valid}}}|} \\ 
  {{R}_{\text{valid}}}=\{(x,y)\ |\ {{I}_{f}}(x,y)>0\ \vee \ I_{{{m}_{\text{ov}}}}^{\text{warp}^*}(x,y)>0\} \\ 
\end{array}
\end{equation}

It reflects the proportion of spatially consistent pixels among valid (non-zero) pixels.

\paragraph{Structural Similarity (SSIM)} 

Local SSIM is computed for each patch $(m,n)$ from $I_f$ and $I_{m_{ov}}^{\text{warp}^*}$:
\begin{equation}
{SSIM}(m,n) = 
\frac{(2\mu_m\mu_n + C_1)(2\sigma_{mn} + C_2)}
{(\mu_m^2 + \mu_n^2 + C_1)(\sigma_m^2 + \sigma_n^2 + C_2)}
\end{equation}
where $\mu$, $\sigma^2$, and $\sigma_{mn}$ denote mean, variance, and covariance. $C_1$ and $C_2$ are stabilization constants. The final SSIM score is the average of local SSIM values over $\Gamma_{{ov}}$.

\paragraph{Mutual Information (MI)}

The mutual information between $I_f$ and $I_{m_{ov}}^{\text{warp}^*}$ over $\Gamma_{{ov}}$ is:
\begin{equation}
{MI}(I_f, I_{m_{ov}}^{\text{warp}^*}) 
= \sum_{i,j} p_{ij} \log \frac{p_{ij}}{p_i p_j}
\end{equation}
where $p_{ij}$ is the joint intensity distribution and $p_i$, $p_j$ are marginal distributions. Higher MI indicates stronger statistical dependence.

\paragraph{Root Mean Squared Error (RMSE)}

The RMSE of overlapping pixels is given by:
\begin{equation}
{RMSE} = \sqrt{\frac{1}{|\Gamma_{{ov}}|} \sum_{(x,y) \in \Gamma_{{ov}}} 
\big( I_f(x,y) - I_{m_{{ov}}}^{\text{warp}^*}(x,y) \big)^2}
\end{equation}
where lower values indicate higher intensity similarity.

\subsubsection{Deformation Field Quality Metrics}

To evaluate the plausibility and physical consistency of the final deformation field $\Phi_{\mathrm{final}}$, we adopt two complementary measures.

\paragraph{Fold Ratio}

A folding occurs when the local mapping is non-invertible, i.e., the Jacobian determinant is negative. The fold ratio is defined as
\begin{equation}
R_{\mathrm{fold}} = \frac{1}{|\Omega|} \sum_{(x,y) \in \Omega} 
\mathbb{I}\big( \det (\nabla \Phi_{\mathrm{final}}(x,y)) < 0 \big)
\end{equation}
where $\Omega$ is the image domain and $\mathbb{I}(\cdot)$ is the indicator function. Lower values indicate smoother, topologically consistent transformations without severe local distortions.

\paragraph{Displacement Standard Deviation}

The displacement at pixel $(x,y)$ is $\| \Phi_{\mathrm{final}}(x,y) - (x,y) \|_2$. The standard deviation of displacement magnitudes is
\begin{equation}
\sigma_{\mathrm{disp}} = \sqrt{ \frac{1}{|\Omega|} \sum_{(x,y) \in \Omega} 
\big( \| \Phi_{\mathrm{final}}(x,y) - (x,y) \|_2 - \bar{d} \big)^2 }
\end{equation}
where $\bar{d}$ is the mean displacement. This metric captures the variability of deformation magnitude; however, it does not directly indicate registration accuracy. Instead, it is used as a supplementary reference to compare the relative deformation intensity across different methods.

\subsection{Experimental settings}

\subsubsection{Experimental Environment Configuration}

The method proposed in this paper was tested on the Ubuntu operating system, using the Python programming language and implemented based on the PyTorch framework. The hardware configuration includes dual Intel® Xeon® Silver 4316 CPUs (2.30 GHz, 20 cores per socket, 80 threads in total), 378 GB of RAM, and a single NVIDIA A100 80GB PCIe GPU for computational acceleration. CUDA version 12.2 and NVIDIA driver version 535.230.02 were used in the experiments.

\subsubsection{Parameter settings}

For superpixel segmentation, we employ the SLIC algorithm with parameters set as follows: number of superpixels $K=30$, compactness factor $\lambda_s = 10$. The input size for the local registration network is fixed to $256\times 256$.

\subsection{Visualization results}

\subsubsection{Global Visualization}

Fig. \ref{fig0} shows the global registration results of the proposed method. Preprocessing significantly enhances intensity uniformity and contrast consistency in both fixed and moving images, providing a solid basis for registration. The initial TPS-based global alignment using sparse control points achieves coarse structural matching of large-scale seafloor features but exhibits local distortions in complex terrain due to smoothness constraints. TPS also reasonably extrapolates non-overlapping areas, serving as a good initialization.

SLIC superpixel segmentation effectively captures seafloor structures, with superpixels aligning well to terrain features and their bounding boxes providing overlapping coverage that enhances robustness in local refinement. The refined global registration improves alignment in overlapping regions by correcting local inconsistencies and better matching feature contours. The final mosaic shows seamless integration without visible artifacts, preserving terrain continuity and confirming the effectiveness of the global-to-local registration strategy.

\subsubsection{Local Visualization}

Fig. \ref{fig1} visualizes the local registration process on representative patches. Locally preprocessed fixed and warped patches (columns 2 and 4) effectively remove striping and speckle noise while preserving seafloor structure clarity.

Compared to initial TPS-warped patches (column 3), the locally registered patches (column 5) show markedly improved alignment with fixed patches (column 1). SynthMorph corrects global-induced distortions, restoring local terrain shape and continuity. Partial overlap between patches allows the same region to appear in multiple contexts, causing slight local variations but enriching the deformation representation for robust global fusion. Displacement fields (column 6) displayed as grid warps reveal fine, smooth, and spatially adaptive local displacements, demonstrating SynthMorph’s capability to model complex deformations without overfitting.

\subsection{Ablation Study}

\begin{figure}[htbp]
	\centering
	\includegraphics[width=1\linewidth]{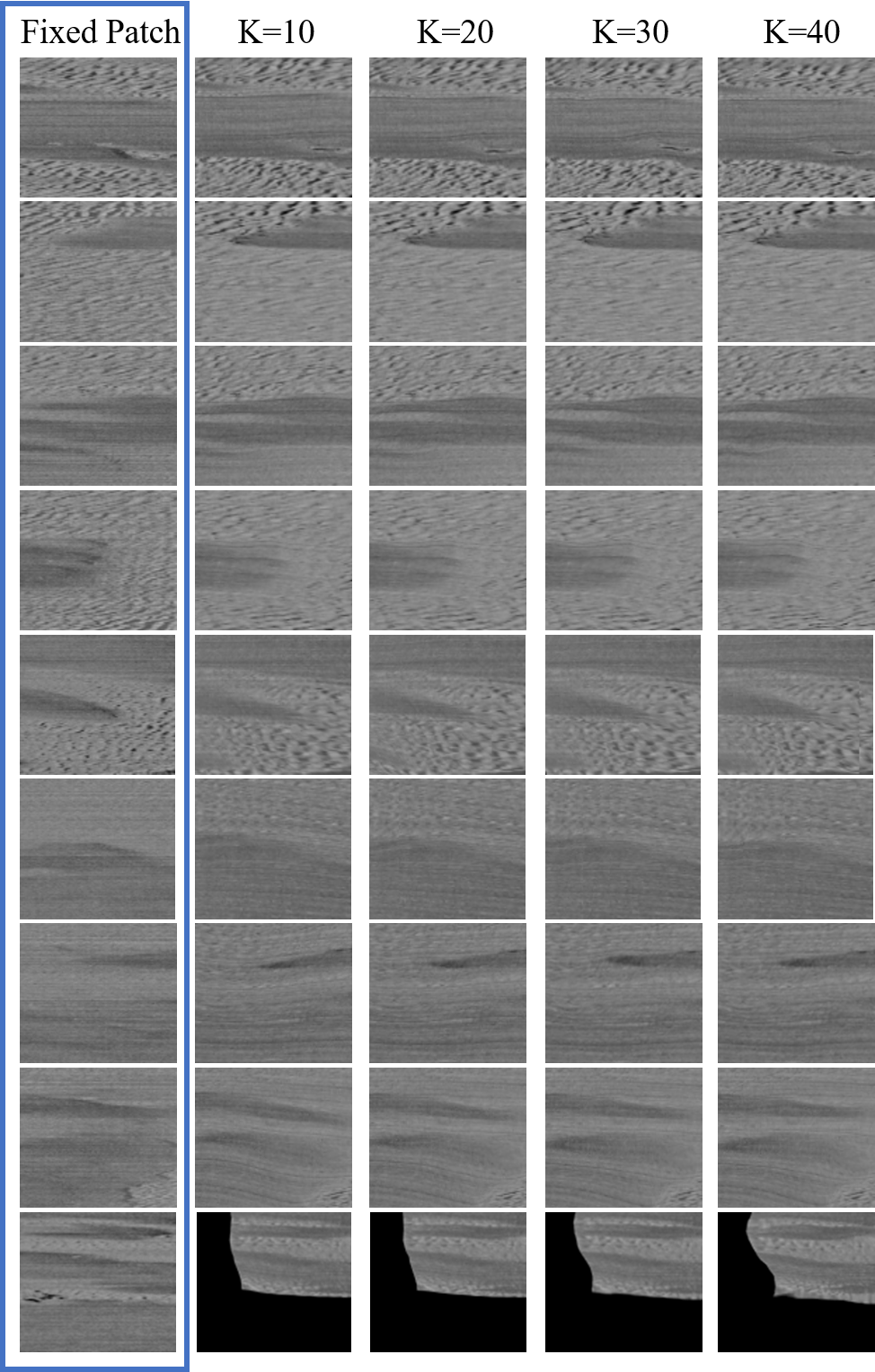}
	\caption{Visual comparison of registration results under different superpixel granularities. Each row corresponds to a distinct overlapping patch region used for local registration. Column 1: Reference fixed patch. Columns 2–5: Locally registered patches using SLIC superpixel counts of $K=10,20,30,40$, respectively. All cases use identical preprocessing settings. Only the number of superpixels used in region segmentation is varied, which directly affects the size and context coverage of each local deformation block.}
	\label{fig3}
\end{figure}	

\begin{figure*}[htbp]
	\centering
	\includegraphics[width=1\linewidth]{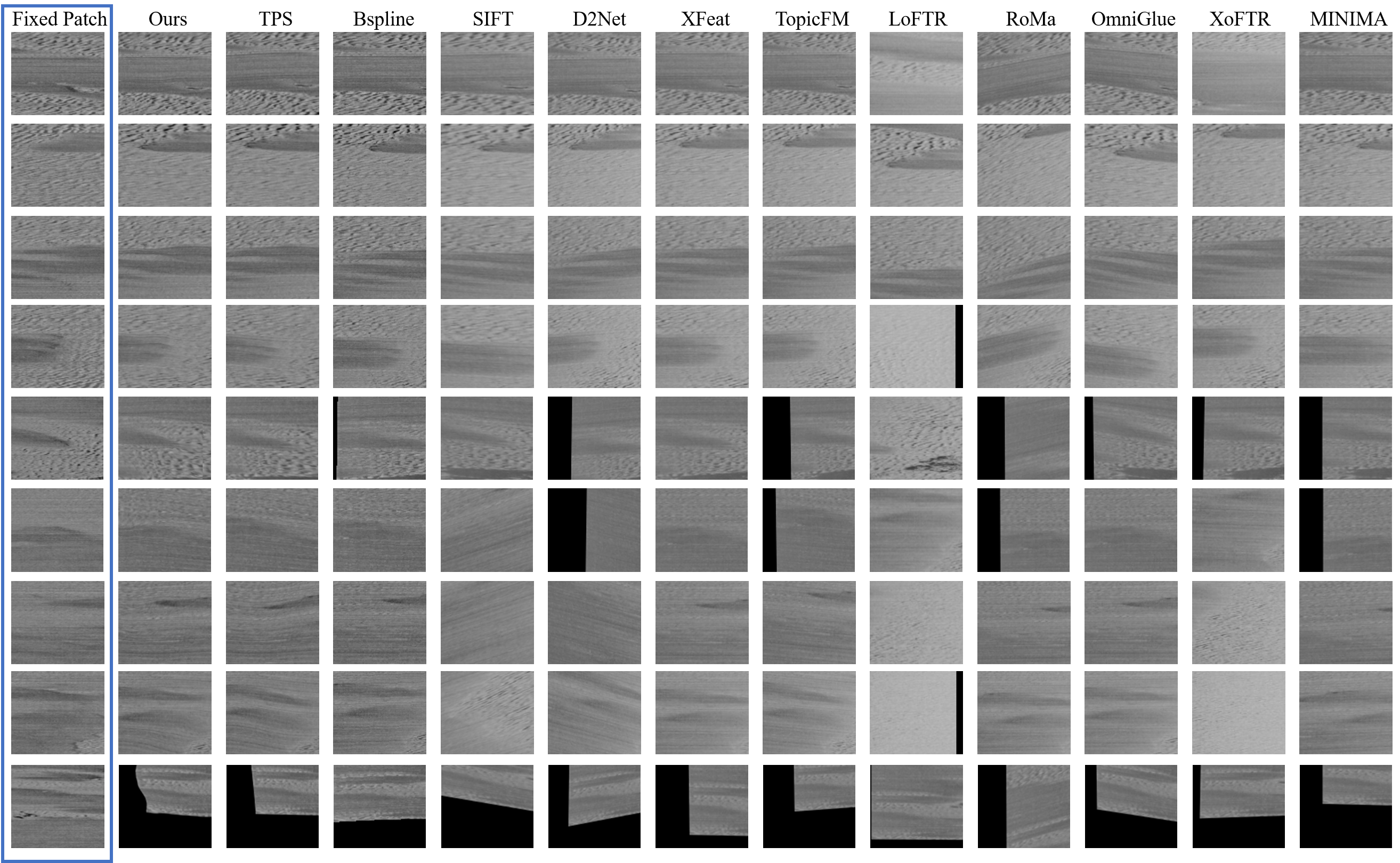}
	\caption{Visual comparison with baseline and state-of-the-art registration methods. Each row corresponds to a different overlapping patch region. From left to right, columns show: The reference fixed patch, our method’s result, results from TPS, B-spline, SIFT, D2Net, XFeat, TopicFM, LoFTR, RoMa, OmniGlue, XoFTR, and MINIMA.}
	\label{fig4}
\end{figure*}

To assess the impact of different components in our pipeline, we perform ablation experiments on two aspects:

\begin{itemize}
    \item \textbf{Preprocessing strategy}: We compare four variants—no preprocessing, patch-level preprocessing only, global-level preprocessing only, and both combined. Global preprocessing homogenizes image intensities between fixed and moving images, while patch-level preprocessing targets stripe artifacts and background noise reduction.
    
    \item \textbf{Superpixel granularity}: We vary the number of superpixels in  segmentation (10, 20, 30, 40) to analyze the effect of region size on local registration performance.
\end{itemize}

\subsubsection{Visual Analysis}

\paragraph{Preprocessing Strategy}

Fig. \ref{fig2} illustrates the impact of different preprocessing schemes on registration. Without preprocessing, registered patches show clear structural distortions and terrain misalignment, indicating poor local alignment on raw inputs. Patch-level preprocessing alone reduces background stripe artifacts but leaves substantial terrain distortion, as it mainly targets non-terrain regions without correcting global intensity differences.

Global-level preprocessing alone greatly enhances registration, yielding patches well aligned with fixed references and improved structural consistency. This highlights the crucial role of global intensity normalization for effective local deformation. Combining both preprocessing steps achieves similarly good alignment, with terrain deformation effectively corrected.

\paragraph{Superpixel Granularity}

Fig. \ref{fig3} shows the effect of superpixel granularity on local registration. Increasing the number of superpixels from 10 to 30 progressively improves terrain alignment, as finer segmentation better captures non-rigid deformations within semantically coherent regions.

However, further increasing superpixels to 40 degrades visual quality, causing over-compensated deformations and unnatural distortions. This likely results from too small regions lacking sufficient structural context, leading to overfitting to minor variations or noise. These findings indicate that while finer granularity enhances local flexibility, excessive fragmentation harms deformation stability. A balance between spatial adaptivity and semantic coherence is essential for robust registration.

\subsubsection{Quantitative Analysis}

Table \ref{tab1} summarizes quantitative results for different preprocessing strategies and superpixel numbers, evaluated by control point error, overlap-based consistency metrics (Overlap Ratio, SSIM, MI, RMSE), and deformation field quality metrics (Fold Ratio, Displacement Standard Deviation).

Global preprocessing consistently yields the largest improvement in registration accuracy. Compared to no preprocessing, global preprocessing reduces control point error (e.g., from 9.37 to 8.35 in TPS-10), increases overlap ratio (from 39.09\% to 52.23\%), and improves SSIM and RMSE. Adding patch-level preprocessing provides marginal gains, indicating limited impact on geometric alignment but potential benefit for visual quality. For superpixel granularity, increasing the number from 10 to 30 steadily enhances performance across all metrics. The best results appear at 30 superpixels, achieving the lowest control point error (5.12), highest overlap ratio (53.41\%), and lowest RMSE. Further increasing to 40 superpixels causes slight degradation, likely due to unstable or over-compensated deformations from excessively fine segmentation.

\subsection{Comparison with existing methods}

\begin{table*}
	\caption{Comparison of Different Registration Methods.}
    \vspace{1mm}
    \small \textbf{Note:}
    ${\bar{e}}$: average control point error; ${R}_{{ov}}$: overlap ratio (\%); $SSIM$, $MI$: structural and statistical similarity; $RMSE$: pixel-wise intensity error; ${R}_{\text{fold}}$: fold ratio (\%); ${\sigma}_{\text{disp}}$: displacement standard deviation.
    
	\label{tab2}
	\centering
	\footnotesize
	\setlength{\tabcolsep}{20pt}
	\renewcommand{\arraystretch}{1.2}
	\begin{tabular}{cccccccc}
		\toprule
		\textbf{Method} & \textbf{${\bar{e}}$(px)} & \textbf{${{R}_{{ov}}}$} & \textbf{$SSIM$} & \textbf{$MI$} & \textbf{$RMSE$} & \textbf{${{R}_{\text{fold}}}$} & \textbf{${{\sigma }_{\text{disp}}}$} \\
		\midrule
        TPS  \cite{A28} & 9.15   & 51.61\% & 0.62 & 0.74 & 15.06   & 27.01\% & 125.95 \\
		B-spline \cite{A29} & 37.85 & 46.12\% & 0.55 & 0.71 & 17.26   & \textbf{0.63\%}  & 106.15 \\    
        SIFT \cite{A12} & 13.07  & 31.18\% & 0.53 & 0.61 & 654.39  & 25.42\% & 97.66  \\
        D2Net \cite{A38} & 29.88  & 42.33\% & \textbf{0.69} & 0.73 & 268.11  & 14.09\% & 168.65 \\
        XFeat \cite{A39} & 13.07  & 33.33\% & 0.65 & \textbf{0.75} & 352.02  & 57.68\% & 96.89  \\
        TopicFM \cite{A40} & 29.10  & 36.74\% & 0.67 & \textbf{0.75} & 321.05  & 46.64\% & 121.69 \\           
        LoFTR \cite{A41} & 220.39 & 21.08\% & 0.45 & 0.39 & 1268.37 & 59.27\% & 5.11   \\
        RoMa \cite{A42} & 53.71  & 39.76\% & \textbf{0.69} & 0.74 & 272.24  & 16.50\% & 103.64 \\
        OmniGlue \cite{A43} & 23.67  & 33.60\% & 0.66 & \textbf{0.75} & 338.67  & 35.30\% & 105.14 \\
        XoFTR \cite{A44} & 209.12 & 22.42\% & 0.47 & 0.55 & 1140.13 & 16.59\% & 27.38  \\
        MINIMA \cite{A45} & 47.84  & 41.45\% & 0.68 & 0.74 & 255.95  & 65.00\%& 115.61 \\      
        \textbf{Ours}  & \textbf{5.12}   & \textbf{53.41\%} & \textbf{0.69} & \textbf{0.75} & \textbf{14.87} & 37.64\% & 126.20 \\
		\bottomrule
	\end{tabular}
\end{table*}

To evaluate the effectiveness of our proposed registration pipeline, we conduct comprehensive comparisons against a diverse set of state-of-the-art methods. These include traditional rigid registration (SIFT), CNN-based rigid matching methods (D2Net \cite{A38}, XFeat \cite{A39}, and TopicFM \cite{A40}), transformer-based matching approaches (LoFTR \cite{A41}, RoMa \cite{A42}, OmniGlue \cite{A43}, XoFTR \cite{A44}, and MINIMA \cite{A45}), as well as non-rigid registration baselines such as TPS and B-spline. All methods are evaluated under consistent data preprocessing and evaluation settings to ensure fairness.

\subsubsection{Visual Analysis}

As shown in Fig. \ref{fig4}, our method consistently produces registration results that are most visually aligned with the reference fixed patches across various scenarios. Compared with other non-rigid baselines (TPS and B-spline), our approach achieves more accurate local alignment and better preserves structural consistency, demonstrating stronger deformation modeling capability. CNN-based rigid methods (D2Net, XFeat, TopicFM) yield partially acceptable results in some regions, but fail to accommodate large non-linear distortions. Traditional keypoint-based methods like SIFT show limited robustness, with noticeable mismatches and deformation artifacts. Transformer-based matchers offer diverse performance: while MINIMA demonstrates better generalization among them, overall results remain less satisfactory than our method, especially in handling fine-scale non-rigid variations. These comparisons highlight the necessity of explicitly modeling non-rigid deformation and validate the effectiveness of our local unsupervised refinement strategy.

\subsubsection{Quantitative Analysis}

As shown in Table \ref{tab2}, our method outperforms all compared approaches across key metrics, achieving the lowest control point error (5.12) and RMSE (14.87), as well as the highest overlap ratio (53.41\%) and SSIM (0.69), demonstrating superior registration accuracy and local consistency.

Traditional non-rigid methods such as B-spline and TPS show moderate accuracy but suffer from drawbacks: TPS attains slightly better control point error than B-spline but exhibits a high fold ratio (27.01\%) and displacement instability (${{\sigma }_{\text{disp}}}= 125.95$), indicating irregular and potentially unreliable deformations. Learning-based methods achieve mixed results. CNN-based matchers (D2Net, XFeat, TopicFM) offer moderate performance and struggle with large deformations common in sonar data. Transformer-based matchers vary widely: MINIMA achieves reasonable SSIM and MI but suffers from the highest fold ratio (65.0\%), reflecting frequent non-invertible warps, while LoFTR and XoFTR produce large errors ($RMSE > 1000$) due to poor generalization in sonar domains.

In contrast, our method balances effective non-rigid modeling and deformation smoothness, maintaining a moderate fold ratio (37.64\%) and displacement standard deviation (126.20). This confirms the robustness of our local refinement strategy in handling complex non-rigid distortions without overfitting.

\section{Conclusion}

This work proposes a hierarchical framework tailored for large-scale side-scan sonar mosaicking, effectively addressing spatially varying geometric distortions. By integrating a global TPS deformation initialized from sparse manual correspondences with a superpixel-guided, patch-wise refinement module based on a pretrained SynthMorph network, our approach balances global structural alignment and fine-grained local adaptation. The superpixel segmentation preserves terrain structure, enabling context-aware local registration that improves deformation accuracy. Extensive experiments validate the superior performance of our method over state-of-the-art rigid, classical non-rigid, and learning-based registration approaches, demonstrating enhanced accuracy, structural consistency, and deformation smoothness on the challenging sonar dataset. Despite these strengths, the method depends on manually selected sparse control points for TPS initialization, which imposes annotation effort and may hinder automation and scalability. Future work will focus on developing automated, reliable sparse correspondence extraction techniques to reduce manual intervention and improve adaptability in diverse sonar environments. In summary, this study advances non-rigid registration for side-scan sonar images by combining classical geometric transformation and unsupervised deep local refinement within a hierarchical design, offering a practical solution for more accurate and reliable large-scale seabed mapping.

\section*{Acknowledgments}

This work was supported in part by the National Natural Science Foundation of China (Grant No. 62171368) and the Science, Technology and Innovation Commission of Shenzhen Municipality (Grant Nos. KJZD20230923115505011, JCYJ20241202124931042), and in part by the Spanish government through projects ASSiST (PID2023-149413OB-I00) and IURBI (CNS2023-144688).

Can Lei also gratefully acknowledges the scholarship support from the China Scholarship Council (CSC).

\bibliographystyle{IEEEtran}
\bibliography{referen}

\end{document}